\documentclass{aa}
\usepackage{graphicx}
\def\hdeux{H$_2$}
\def\c2{C$_2$}
\def\chp{CH$^+$}
\def\lya{Ly$\alpha$}
\def\lyb{Ly$\beta$}
\def\lyc{Ly$\gamma$}
\def\lyd{Ly$\delta$}
\def\cmd{cm$^{-2}$}
\def\cmt{cm$^{-3}$}
\def\kms{km s$^{-1}$}
\def\CHl{CH~$\lambda$4300}
\begin{document}
   \title{A far UV study of interstellar gas towards HD~34078: high
excitation \hdeux\ and small scale structure\thanks{Based on observations
performed by the FUSE mission and at the CFHT telescope}}

\titlerunning{A far UV study of interstellar gas towards HD34078}

   \author{P. Boiss\'e\inst{1, 2}
          \and
          F. Le Petit\inst{3, 4}
          \and
          E. Rollinde\inst{1, 5}
          \and
          E. Roueff\inst{3}
          \and
          G.~Pineau des For\^ets\inst{6, 3}
          \and
          B-G~Andersson\inst{7}
          \and
          C.~Gry\inst{8, 9}
	  \and
          P.~Felenbok\inst{10}
          }

   \offprints{P. Boiss\'e}

   \institute{Institut d'Astrophysique de Paris, Paris, France
         \and
            LERMA/ENS, France
         \and
             LUTH, Observatoire de Paris-Meudon, Meudon, France
	\and
             Onsala Space Observatory, 439 92 Onsala, Sweden
         \and
            Institute of Astronomy, Cambridge, UK
         \and
            IAS, Universit\'e d'Orsay, Orsay, France 
	 \and
	    Johns Hopkins University, Baltimore, USA
	 \and
            ESA, Vilspa, Spain
         \and
            Laboratoire d'Astrophysique de Marseille, Marseille, France       
         \and
            LESIA, Observatoire de Paris-Meudon, Meudon, France
             }

   \date{Received January 27; accepted July 7, 2004}

\abstract{
To investigate the presence of small scale
structure in the spatial distribution of H$_2$ molecules we
have undertaken repeated FUSE UV observations of the runaway
O9.5V star, HD~34078. In this paper we present five
spectra obtained between January 2000 and October 2002.
These observations reveal an unexpectedly large amount of highly excited
H$_2$. Column densities for H$_2$ levels from (v = 0, J = 0) up
to (v = 0, J = 11) and for several v = 1 and v = 2 levels are determined.\\
These results are interpreted in the frame of a model involving 
essentially two components:  i) a foreground cloud (unaffected by 
HD~34078) responsible for the \hdeux\ (J = 0, 1), CI, CH, \chp\ and CO 
absorptions; 
ii) a dense layer of gas (n $\simeq 10^4$ \cmt) close to the O star
and strongly illuminated by its UV flux which accounts 
for the presence of highly excited \hdeux. Our model
successfully reproduces the \hdeux\ excitation, the CI fine-structure 
level populations as well as the CH, \chp\ and CO column densities.\\
We also examine the time variability of \hdeux\ absorption lines tracing each
of these two components. From the stability of the J = 0, 1 and 2 damped
\hdeux\ profiles we infer a 3$\sigma$ upper limit on column density 
variations $\Delta$N(\hdeux)/N(\hdeux) of 5\% over scales ranging from 5 to 
50 AU. This result clearly rules out any pronounced ubiquitous small scale 
{\it density} structure of the kind apparently seen in HI. The lines from 
highly excited gas are also quite stable (equivalent to $\Delta$N/N
$\le$ 30 \%) indicating i) that the ambient gas through which HD~34078 
is moving is relatively uniform and ii) that the gas flow along the 
shocked layer is not subject to marked instabilities.\\
}
\maketitle
%

\section{Introduction}

Studies of the molecular phase of the interstellar medium are severely
hampered by the fact that the most abundant species, H$_2$, cannot be
observed from the ground. "Tracers", like CH, CO and its isotopomers are
often used as surrogates; however, because of selective photodissociation
near the illuminated boundaries of clouds, depletion onto grains in their
depths 
or, simply, time evolution under the coupled effect of internal
dynamics and chemistry, no tight relation is expected between the {\it
local} 
abundance of H$_2$ and those of minor species. In other words, the
widely used notion of "standard" ratios can hardly be justified,
even if, when integrated along the line of sight, the amounts
of various species appear relatively well correlated.

The above limitation is especially severe when we deal with small
scale structure studies. Within translucent molecular
clouds, H$_2$CO, OH and HCO$^+$ molecules have been found to display
AU-scale structure (Moore \& Marscher \cite{moore95}, Liszt \& 
Lucas \cite{liszt00}). In contrast, the distribution of dust grains 
is quite smooth
at small scales, both in translucent and dense material (Thoraval et
al. \cite{thoraval96}, \cite{thoraval97}, \cite{thoraval99}; 
Lada et al. \cite{lada99}; Alves et al. \cite{alves01}). If H$_2$
were to display
marked AU-scale structure, as minor molecular species apparently do,
then the smooth behavior of dust grains could be understood as a result of
their large inertia (Thoraval et al. \cite{thoraval99}). However, it could
also be that the H$_2$ distribution is smooth at small scales with no
or little density structure (like dust grains) and that "chemical"
inhomogeneities are
present. Clearly, direct investigations of the distribution of H$_2$
itself are needed to clarify the above issues and determine whether
the structure seen for minor species is due to fluctuations of their
local abundance or rather reflects real {\it density} structure.

We have thus taken advantage of the unique opportunity offered by
the successful launch of FUSE and proposed a GI program which aims
at probing the distribution of H$_2$ over spatial scales ranging from
about 1 to 100 AU. A bright O9.5V star, HD~34078 (AE Aur) has been selected
for repeated observations and, thanks to its large proper motion
(Moffat et al. \cite{moffat98}), 
information on the structure of the foreground medium can be obtained 
within a relatively small time interval. Indeed, at a distance of
530 pc the implied transverse velocity is 103 \kms\ or 22 AU/yr. Its
color excess, E(B-V) = 0.52, is such that a significant fraction of the
foreground gas should be in molecular form. It is also of interest that the
velocity structure of the molecular gas along this line of sight is
quite simple, with only one major component that has b $\approx$ 3 \kms\ 
(Allen \cite{allen94}, Rollinde et al. \cite{emmanuel}). 
In the following, we shall adopt a distance of 380 pc for this
molecular material (Brown et al. \cite{brown95}), implying a
drift of the line of sight through the cloud at a velocity of 17 AU/yr.
Note finally that HD~34078 has been the
subject of several detailed studies in the past decade that have brought
results on a number of molecular species including CH, CH$^+$, CN,
C$_2$ (Federman et al. \cite{federman94}, Herbig \cite{herbig99}). The 
latter can be helpful for modelling and the past measurements provides
 information on structure over  
larger scales by comparison with present-day observations. In
particular, a series of observations of the blue CH and \chp\
lines has been performed since 1999 in parallel with this FUSE program
(Rollinde et al. \cite{emmanuel}). Results indicate a long-term 
increase of N(CH) over the past 10 years while N(\chp) remained stable 
or decreased. Correlated short-term ($\simeq$ 1 yr) CH and \chp\
fluctuations might also be present.

\begin{table}[h!]
\label{tab.param}
\caption{Properties of the line of sight. Right side of the chart: extinction curve coefficients.}
\begin{tabular}{lc|cc}
\hline
\hline
Spectral Type$^{(1)}$    & O9.5Ve       & \multicolumn{2}{c}{Ext. curve$^{(4)}$}\\
$\alpha^{(1)}$ (J2000)   & 05 16 18.15  & c$_1$ & 0.473      \\
$\delta^{(1)}$ (J2000)   & +34 18 44.3  & c$_2$ & 0.571      \\
m$_\textrm{v}^{(1)}$     & 5.998        & c$_3$ & 4.152      \\
parallax$^{(1)}$ (mas)   & 2.24         & c$_4$ & 0.52       \\
E(B-V)$^{(2)}$           & 0.52         & $\gamma$ & 1.087   \\
R$_\textrm{v}$$^{(3)}$   & 3.42         & x$_0$ & 4.589      \\
\hline
\end{tabular}
\newline
{\small 
(1) Simbad database\\
(2) Crawford (\cite{crawford})\\
(3) Patriarchi et al. (\cite{patriarchi01})\\
(4) Fitzpatrick and Massa (\cite{fitzpatrick})
}
\end{table}

The first FUSE spectrum obtained revealed large amounts of
high excitation H$_2$. This is an unexpected finding in comparison
with observations of most other lines of sight (see however 
Meyer et al. \cite{meyer01} for HD~37903 and Federman
et al. \cite{federman95}).
As HD~34078 is a high velocity star, it is likely to be far away from
its parent cloud and should hence not be associated with molecular 
material. This result has important
implications for the identification of physical processes responsible
for the unusual excitation but could be a difficulty for our structure
study. Indeed, the presence of highly excited H$_2$ raises the
possibility of a significant contribution to the observed absorptions
from gas associated with the star itself, which could be subject to
variability of a different nature than the one due to structure in the
foreground gas. The first objective of this paper therefore becomes to
establish a reasonable scenario for the origin of the observed
absorption, based on detailed modelling of the HD~34078 FUSE spectra. 
Secondly, we present a study of the variability of the 
\hdeux\ lines based on five spectra spanning a period of 2.7 years.

In Sect. 2, we describe the FUSE observations and data
analysis. Complementary OH observations performed at the CFHT are
presented in Sect. 3. Results 
on the gas properties are discussed in Sect. 4 with their
implications, as inferred from modelling of the physical conditions
in the absorbing material. We next investigate \hdeux\ line variations
(Sect. 5).
Preliminary analyses of the first three FUSE spectra have been presented by
Le Petit et al. (\cite{lepetit00}, \cite{lepetit01}) and Boiss\'e et al. (\cite{patrick01}).

\section{The FUSE data}

\subsection{Observations and data reduction}

The first FUSE spectrum was obtained on 2000 January 23;
it consisted of eleven exposures with a total integration
time of 7145~s . Four other spectra were acquired in comparable
conditions on 2000 October 30, 2001 February 16, 2001 October 15 
and 2002 October 10 (cycle 1 and 2 programs). The data were taken in
histogram (HIST) mode. 
In this paper we rely mostly on the  
first spectrum for the study of the intervening gas properties,
but occasionally we use the other four, either to
secure detections of faint lines or to improve the accuracy
of column density estimates. The search for variable \hdeux\ absorption
has been performed mainly by comparing the 1st and 5th spectra,
because i) they turn out to be of very similar "quality" (resolution
and S/N), ii) they sample the largest time interval (i.e. largest
spatial scales, over which changes might be easier to detect in the
event of a regular variation). For absorption lines that are suspected
to vary in time, measurements are performed on the first spectrum.

Standard processing using the FUSE pipeline software was applied
to perform the wavelength and flux calibration (for details on the
FUSE mission, see Moos et al. \cite{moos00} and Sahnow et al. \cite{sahnow00}). The
analysis presented here is based mostly on the LiF1A and LiF1B
spectra, which display the highest S/N ratio.

The data were reduced in a
standard way using MIDAS, the ESO data reduction software package.
Comparison of individual exposures occasionally revealed variations
in absorption line positions from one exposure to
another. Nevertheless, these remain quite small and the average
spectrum obtained after applying the appropriate shifts does not
differ significantly from the raw average. The original pixel
corresponds to 0.005 \AA\ but, given the resolution provided by FUSE
(about 20 \kms) the spectra were binned to 0.015 \AA.
In the co-added spectrum,
the S/N ratio reaches 30 per 15 m\AA\ pixel which
translates into a 3 $\sigma$ limit on the equivalent width (W) 
as low as 4 m\AA.

\subsection{Line identification and column density determinations}
Line identification has been performed using the list provided by
Morton (\cite{morton00}) for atoms and ions. For \hdeux\ transitions we use
the data from Abgrall et al. (1993ab).  Originally, only lines from
v = 0 with J = 0, 1, ....9 levels were considered. As it turned out
that lines from higher excited levels were present in our spectra,
the table was complemented by adding transitions from v =
0, J = 10, 11, 12, 13;  v = 1,  J = 0, ..., 9 and v = 2, 3, 4,
 J = 0, 1, 2. 

\begin{figure*}
\centerline{
 \includegraphics[width=13cm,angle=-90]{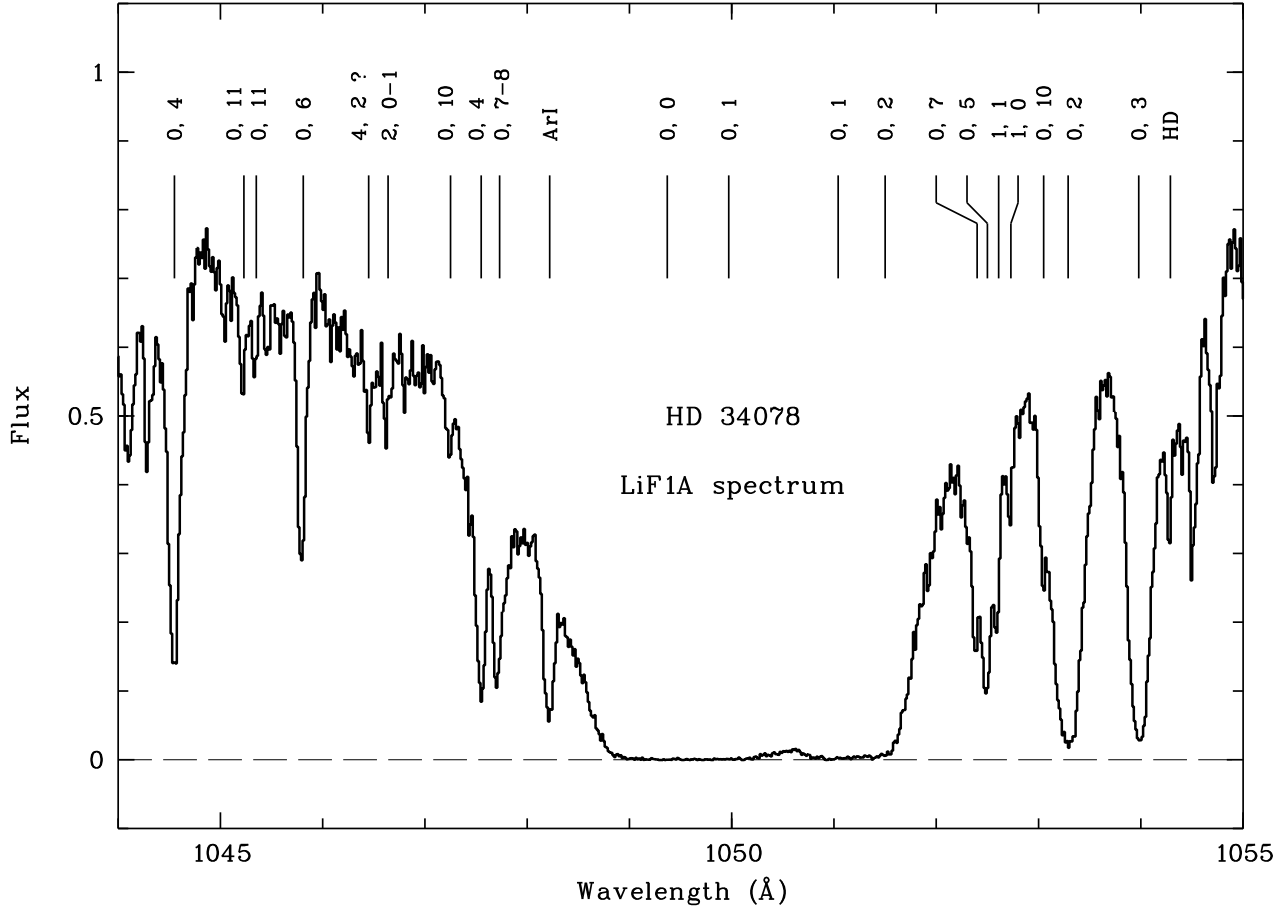}
}
\caption{Portion of the January 2000 LiF1A spectrum in units of
10$^{-11}$ erg s$^{-1}$ \cmd \AA$^{-1}$ near 
strong \hdeux\ (J = 0, 1, 2) lines from the (4-0) Lyman band. v and J values
for the lower level of each \hdeux\ transition involved are given.
The features indicated illustrate the variety of excited \hdeux\ levels 
from which absortion is detected. The identification of the feature 
noted ``4, 2 ?'' is uncertain (see text in Sect. 2.3).}
\label{f:spectre}
\end{figure*}

To perform the line identification, we proceed in two steps.
We first measure the position of all prominent absorption lines in
the spectrum and assign one (or several) transitions to each of them.
Next, we search in a systematic way for all strong transitions
expected in our range and examine whether all 
detections/non-detections and W values for any given species 
(or particular level
of a species) are mutually consistent. This step is a crucial one
here since, given the high number density of absorption lines in
the spectrum, blends are numerous and the continuum level is often
ill-defined (we estimate that in the 1040 - 1060 \AA\ interval 
for instance, the
fraction of the spectrum free of absorption is no larger than 25\%;
this fraction becomes close to zero above 1100 \AA). As a result of
this crowding, some lines may appear either too strong in the event
of an unresolved blend or too faint if adjacent absorption leads
to an underestimate of the continuum level. When an unresolved blend
or ill-defined continuum is suspected, the corresponding feature
is not used for column density determination. A portion of the LiF1A
spectrum is shown in  Fig. \ref{f:spectre} where several high excitation 
\hdeux\ lines are identified.

\begin{figure}
\centerline{\includegraphics[width=6.8cm,angle=-90]{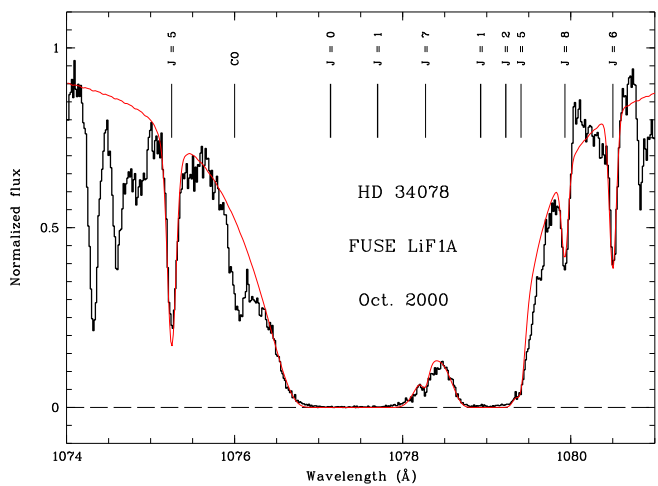}}
\caption{Portion of the October 2000 LiF1A FUSE spectrum comprising
the strong \hdeux\ (2-0) Lyman band around 1078 \AA\ (red line). 
The fit obtained when including the nine \hdeux\ lines indicated 
(v, J values for the lower level are given) is superimposed 
(thin line); thick tick marks indicate 
the four strong damped \hdeux\ lines from J = 0, 1, 2. 
The absorption near 1076 \AA\ is from CO.}
\label{f:spectre_fit}
\end{figure}

Except for damped H$_2$ lines, most features are not resolved
(recall that b $\simeq$ 3 \kms\ for molecular gas towards HD~34078).
Further, the line spread function is not accurately known and may
vary across the spectrum; we then often rely on curve of growth
analysis to derive column densities. However, line fitting is required
to deal with partially resolved blends and damped H$_2$ lines. To this
purpose, we use the routine OWENS developed by Martin Lemoine and
the LYMAN context in MIDAS. In a first step, the spectrum is
shifted to zero velocity using narrow features located blueward
and redward of the absorption considered. When fitting damped \hdeux\
profiles, the continuum is interpolated using low order polynomials.
Good fits are easily obtained as shown in Fig. \ref{f:spectre_fit}.
For low J values, several H$_2$ systems are available that can be
used to get independent column density estimates (hereafter N), which 
helps to further reduce the uncertainties. 

\begin{table*}[h!]
\begin{center}
\label{nh2}
\caption{Measured column densities for various \hdeux\ levels (best
estimate and 3 $\sigma$ lower and upper bounds) and predictions from
our model (see text). The number of lines effectively used in the
determination of N(H$_2$, v, J) is given as M$_{l}$}
\begin{tabular}{rrrrr|ccc|cccc} 
\cline{1-12}
    &    &    &       &    & \multicolumn{3}{c}{Observations} & 
\multicolumn{4}{c}{Modelling} \\
 \cline{1-12}
v   & J  & g  & Energy& M$_{l}$& N(H$_2$, v, J)& Lower & Upper & 
Transluscent& C-shock   & hot PDR & Total        \\
    &    &    & (K)   &      & (cm$^{-2}$) & limit &  limit &  Cloud    
&           &         &     \\
 \hline
  0 &  0 &  1 &     0 &  5 & 3.2(20) & 3.0(20) & 3.6(20) & 3.2(20) & 8.6(17)&  
2.1(18) & 3.2(20)\\
  0 &  1 &  9 &   170 & 10 & 3.2(20) & 2.9(20) & 3.5(20) & 3.2(20) & 3.1(18)& 
1.4(19) & 3.4(20)\\
  0 &  2 &  5 &   510 &  3 & 1.8(19) & 1.4(19) & 2.3(19) & 2.0(18) & 4.4(17)& 
4.0(18) & 6.4(18)\\
  0 &  3 & 21 &  1015 &  5 & 6.2(18) & 5.5(18) & 8.0(18) & 6.1(15) & 6.5(17)& 
6.0(18) & 6.7(18)\\
  0 &  4 &  9 &  1682 &  4 & 7.1(17) & 2.5(17) & 1.1(18) & 2.0(14) & 7.0(16)& 
5.5(17) & 6.2(17)\\
  0 &  5 & 33 &  2504 &  4 & 3.3(17) & 7.8(16) & 4.5(17) & 7.6(13) & 4.7(16)& 
1.9(17) & 2.4(17)\\
  0 &  6 & 13 &  3474 &  5 & 4.0(15) & 1.8(15) & 2.6(16) & 8.4(12) & 2.8(15)& 
3.6(15) & 6.4(15)\\
  0 &  7 & 45 &  4586 &  1 & 2.5(15) & 5.4(14) & 9.0(15) & 7.5(12) & 1.4(15)& 
2.0(15) & 3.4(15)\\
  0 &  8 & 17 &  5830 &  5 & 6.0(14) & 4.5(14) & 1.4(15) & 1.1(12) & 9.1(13)& 
2.5(14) & 3.4(14)\\
  1 &  0 &  1 &  5987 &  2 & 2.2(13) & 2.0(13) & 5.0(13) & 2.7(11) & 1.4(12)& 
2.9(13) & 3.1(13)\\
  1 &  1 &  9 &  6150 &  2 & 2.1(14) & 1.7(14) & 5.0(14) & 7.7(11) & 1.2(13)& 
1.4(14) & 1.5(14)\\
  1 &  2 &  5 &  6472 &  3 & 8.3(13) & 7.0(13) & 2.0(14) & 6.8(11) & 5.1(12)& 
9.8(13) & 1.0(14)\\
  1 &  3 & 21 &  6952 &  4 & 1.7(14) & 1.2(14) & 2.5(14) & 6.9(11) & 1.5(13)& 
1.8(14) & 2.0(14)\\
  0 &  9 & 57 &  7197 &  8 & 6.0(14) & 2.0(14) & 8.0(14) & 1.5(12) & 4.6(13)& 
4.5(14) & 5.0(14)\\
  1 &  4 &  9 &  7585 &  4 & 7.0(13) & 6.0(13) & 9.0(13) & 3.0(11) & 4.1(12)& 
8.0(13) & 8.4(13)\\
  1 &  5 & 33 &  8365 &  3 & 1.8(14) & 1.5(14) & 2.3(14) & 2.4(11) & 7.9(12)& 
1.2(14) & 1.3(14)\\
  0 & 10 & 21 &  8677 &  2 & 4.0(13) & 2.5(13) & 6.0(13) & 2.5(11) & 3.6(12)& 
8.1(13) & 8.5(13)\\
  0 & 11 & 69 & 10262 &  3 & 4.1(13) & 2.1(13) & 6.1(13) & 4.1(11) & 1.8(12)& 
1.5(14) & 1.5(14)\\
  2 &  0 &  1 & 11636 &  1 &------  & ------  & 3.0(13) & 1.1(11) & 1.9(10)& 
1.3(13) & 1.3(13)\\
  2 &  1 &  9 & 11789 &  1 &------  & ------  & 3.0(14) & 3.0(11) & 1.6(11)& 
6.1(13) & 6.1(13)\\
  0 & 12 & 25 & 11940 &  2 &------  & ------  & 2.4(13) & 7.5(10) & 1.5(11)& 
3.1(13) & 3.1(13)\\
  0 & 13 & 81 & 13703 &  1 &------  & ------  & 2.7(13) & 1.4(11) &  
------&6.2(13) & 6.2(13)\\
 \hline
\end{tabular}
 \end{center}
 \end{table*}

\begin{figure}
\centerline{\includegraphics[width=6.4cm,angle=-90]{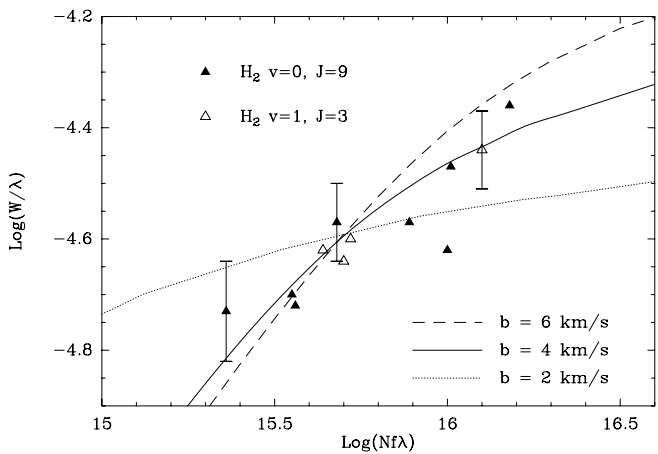}}
\caption{Curve of growth for \hdeux\ transitions from the v = 0,
J = 9 and the v = 1, J = 3 levels (filled and empty triangles
respectively). The horizontal scale is appropriate for  b = 4
\kms; the slopes of the curve of growth corresponding to b = 2
and b = 6 \kms\ are indicated (dotted and dashed line, respectively).
A few representative 1 $\sigma$ error bars are given.}
\label{f:cog_J9}
\end{figure}

Damped lines from low energy \hdeux\ levels provide a measurement of
N which is independent of the velocity distribution of the gas;
for a b parameter of about 3 \kms\ (as measured for CH), we
find that this is true up to J = 3. The situation is less comfortable
for lines of intermediate optical depth. Indeed, difficulties in
defining the continuum or due to blends are
such that for most H$_2$ levels we cannot build a curve of
growth appropriate for deriving both the b parameter and N. We have
to assume some a priori b value. 
To investigate the possibility of a variation in b with excitation 
energy, we searched for
highly excited levels (J $>$ 5) for which our data can be used to constrain
the b value. Thus, for the (v = 0, J = 9) level, we get reliable
measurements for 8 transitions with log($\lambda$ f) values ranging
from 0.75 to 1.57 (cf. Fig. \ref{f:cog_J9}). We also include three
measurements from 
the (v = 1, J = 3) level which is close to the previous one in
the energy scale. As can be seen in Fig. \ref{f:cog_J9}, both levels
give consistent constraints on the curve of growth. From these, we
estimate that a b value of about 4 \kms\ is consistent
with the data and find that values below 2 and above 6 \kms\ are
excluded. We therefore find no evidence that the velocity distribution 
of the high excitation \hdeux\ differs from that of CH, which 
presumably reflects that of the bulk of \hdeux. Note that 
Meyer et al. (\cite{meyer01} reach a similar conclusion for highly excited
\hdeux\ detected in front of HD~37903 (the b value being 1.8 \kms\ 
in this case). In practice we shall adopt b = 4 \kms\ for J $>$ 3.
 
Optically thin \hdeux\ lines can also be of interest for deriving N
without any assumption on the velocity distribution, and for providing
reliable constraints on the upper end of the excitation diagram. For
b = 4 \kms we estimate that a line around 1000 \AA\ and with an
equivalent width of less than 8 m\AA\ can be considered to be
optically thin (the N values derived either from the exact N(W)
relation or the linear approximation differ by less than 10\%).
This is significantly above our detection limit and we do detect
\hdeux\ features of this type from the v = 0, J = 10 and J = 11 or
some v = 1 levels. Non-detections in regions with good S/N can
also be used to derive upper bounds on N (e.g. for \hdeux\ v=0,
J = 12 and 13).

Table \ref{nh2} summarizes the results obained for N(\hdeux, v, J) 
and gives the
best values together with 3 $\sigma$ lower and upper bounds. The
latter have been estimated in a conservative way. When different lines
from the same level provide discordant results for N, we adopt the
best compromise by weighting various estimates according to the
presence of suspected blends and quality of the data. The number 
of lines effectively used in the determination of N is given
in Table \ref{nh2} as M$_{l}$. For lines which
are neither in the damped or optically thin regime, 
an uncertainty on b of $\pm$ 1 \kms\ was considered. 

\subsection{Comments on specific \hdeux\ levels or species}
In the following, we give some details concerning the column density
determination for transitions:

$\bullet$ H$_2$ J = 0, 1, 2 :
For the absorption from these levels, the continuum level and
shape across the broad profiles is difficult to define. To get an
estimate of the unabsorbed spectrum, we have artificially removed all narrow
features other than those from H$_2$ J = 0, 1, 2. We then simultaneously
fit the latter using OWENS; a polynomial form of order 3 has been adopted
for the continuum. Fortunately, N(J = 0, 1, 2) is well
constrained by the profile near the bottom of the line and weakly
depends on the exact shape of the adopted continuum.

$\bullet$ H$_2$ J = 3, 4, ..., 9 : for these levels, the number 
of lines available range from 1 (J=7) to 10 (J=9).   

$\bullet$ H$_2$ v = 0, J = 11 :
This is the highest v = 0 transition detected. Two features are seen
at 1045.22 and 1045.35 \AA\ with W = 10 $\pm$ 2 m\AA\ and
W = 7 $\pm$ 2 m\AA.

$\bullet$ H$_2$ v = 0, J = 12 and 13 :
Lines at 1057.62 and 1072.73 \AA\ from J = 12 are expected in good
"windows"; from their absence we get the upper limit 
N $<$ 2.4 10$^{13}$ cm$^{-2}$. 
The strongest transition in our range from J = 13 is expected at
1040.70 \AA\ (f = 0.0237) and is marginally present (2 $\sigma$
detection). We infer the upper limit: N $<$ 2.7 10$^{13}$ cm$^{-2}$.

$\bullet$ H$_2$ v = 1 : several features are seen from each
of the J = 0 to 5 levels. Therefore, their identification is
secure. In contrast, no feature from the (v = 1, J = 6, 7, 8) levels
could be measured due to blending with strong absorptions.

$\bullet$ H$_2$ v = 2, 3, 4 :
A blend of the 1046.62 and 1046.66 \AA\ lines (from v = 2 , J = 1 and 0
respectively) is marginally detected with W = 12 m\AA\ in the 1st, 4th
and 5th spectra, which have the best resolution and S/N in this region.
Unfortunately, we do not find other lines suitable for the
determination of N(v = 2 , J = 0) and N(v = 2 , J = 1) separately.
  
Regarding v = 3 levels, several lines are suspected to be present but
have a low significance level. For instance, a line from (v = 3, J =
1) is seen in the 1st and 4th spectra at 1116.405 \AA\ with W = 3
$\pm$ 1.5 m\AA. The corresponding column density is N = 6
10$^{12}$cm$^{-2}$. However, most of these lines are expected at 
$\lambda$ $>$ 1100 \AA\ where the crowding is quite severe, and we 
prefer to retain only upper limits for these levels.

In an earlier report (Boiss\'e et al. \cite{patrick01}), we assigned 
the 1046.45 \AA\ feature to the v = 4, J = 2 level (line 5 in Fig. 1). 
This line appears
clearly only in the first spectrum and given i) the small f value
(6 10$^{-3}$), ii) absence of other detections from the same level 
and iii) absence of time variations of clearly detected highly excited
\hdeux\ features, this identification appears questionable. 
Only spectra from the HST/STIS would allow us to unambiguously
quantify the amount of H$_2$ in levels with excitation energies 
above 10$^4$ K.

$\bullet$ HI :
In our data, the \lyb\ line is present although heavily blended with
\hdeux\ J = 0, 1, 2, 5 absorptions near 1025 \AA\ (\lyc\ and \lyd\ can also be
seen but the S/N ratio degrades rapidly at lower wavelengths and
blends are numerous so we use only \lyb\ to constrain
N(HI)). Thanks to the large width of this \lyb-\hdeux\ absorption,
airglow emission does not affect its profile (emission occurs only in the
central part of the trough, where the flux is zero; Fig. \ref{f:HI}).
To fit the spectrum in this region, we adopt the values for
N(H$_2$, v=0, J) inferred from other H$_2$ systems and vary N(HI). The
shape of the red edge of the broad 1025 \AA\ trough (around 1027 \AA) 
is primarily due to \lyb\ (see Fig. \ref{f:HI}) from which we 
estimate N(HI) = 2.7 $\times$ 
10$^{21}$ \cmd, a value significantly larger than that derived by Mc
Lachlan \& Nandy \cite{maclac} (N(HI) = 1.75 $\times$ 10$^{21}$ \cmd) 
from IUE \lya\ data taken in 1979. We come back to N(HI) variations in 
Sect 5.4.

$\bullet$ HD, J = 0 :
Several lines are detected (near 1007, 1011, 1021, 1031, 1042,
1054, 1066 \AA) but unfortunately, most of them are either blended,
located in regions of poor S/N or have an ill-defined continuum.
For instance, the strong 1021.456 \AA\ line is expected to be blended
with an SII feature with an unknown  oscillator strength. The 1031 \AA\
feature is contaminated by an \hdeux\ (v = 1, J = 3) line  for
which N is reasonably well determined; OVI$\lambda$1031 might also
contribute, but this cannot be assessed since the OVI$\lambda$1037 line 
coincides with a damped H$_2$ line. The 1054~\AA\ line appears much too faint
with respect to the others (W $\approx$ 15 m\AA); probably, this is
due to an ill-defined continuum caused by extra absorption adjacent
to this line. In contrast, the 1066.27 \AA\ feature is too strong
with respect to the 1031.91 \AA\ one. It is thus difficult to
infer a value for N(HD, J = 0) consistent with all lines
detected. Moreover, N values are strongly dependent on the velocity
distribution, which might differ from that of CH. We therefore cannot
do better than give an order of magnitude estimate: N(HD, J = 0) 
$\approx 10^{15-16}$ \cmd .

$\bullet$ HD J = 1 :
Lines from excited HD have already been detected in the
Copernicus spectrum of $\zeta$ Oph (Wright \& Morton 
\cite{wri79}). The strongest feature mentioned
by these authors, at 1021.916 \AA, is apparently present in our five
spectra, although blended with other weak lines. We estimate $W
= 8 \pm 2$ m\AA. Other lines of comparable strength are expected from
HD J = 1; at the position of the 1054.722 \AA\
transition, a strong line with W $\approx$ 35 m\AA\ is present but it
is much too strong in comparison to the previous feature and is likely
of stellar origin. Finally, around 1043.288 \AA\ (on the blue wing of
a \hdeux\ J = 3 line) the spectra display large variations which are of
instrumental or stellar origin and the search is inconclusive.
The tentative detection of the 1021.916 \AA\ line implies
N(HD, J = 1) = 5.4 $\pm 1.3 \times 10^{13}$ \cmd .

$\bullet$ CI : 
Numerous lines from CI and its fine structure excited
levels are detected which can be used to measure N(CI), 
N(CI$^{*}$) and N(CI$^{**}$) and get constraints on the density, n. 
This key parameter has been estimated by Federman et al.
(\cite{federman94}) from the analysis of \c2\ excitation.
 
Several combinations of lines from three multiplets were considered 
between 1129 and 1160 \AA; spectral regions containing lines with
unknown f values were excluded. Various attempts were made to fit line
profiles using OWENS; we successively considered either unblended or
unsaturated features alone or all of them together. The instrumental
LSF was allowed to vary slightly across the spectrum and b values 
between 3 and 6 \kms\ were considered. The results are the following:

- N(CI) = 9.4 (+ 7.0, - 5.8) 10$^{15}$ \cmd,

- N(CI$^{*}$) = 5.8 (+ 0.0, - 4.2) 10$^{15}$ \cmd,

- N(CI$^{**}$) = 2.2 (+ 1.8, - 1.1) 10$^{15}$ \cmd.

These estimates and uncertainties represent the best compromise we
find given our limited knowledge of the LSF and velocity distribution.

\section{The CFHT data}
\begin{figure}
\centerline{\includegraphics[width=6.4cm,angle=-90]{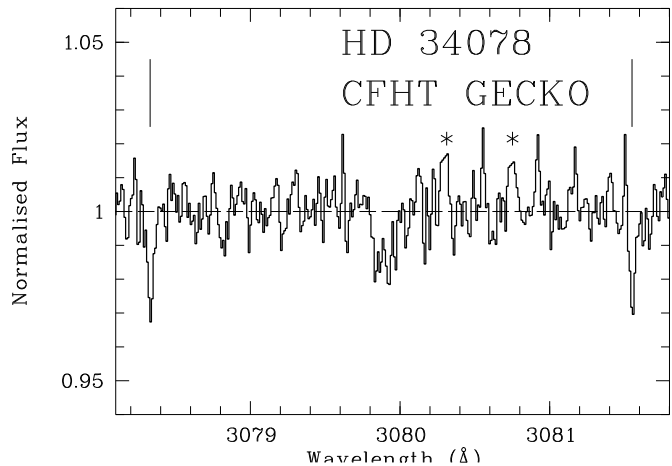}}
\caption{Portion of the normalised CFHT spectrum comprising two 
OH absorption lines (indicated by tick marks). The broad absorption 
around 3079.9~\AA\ is due to a detector defect. High intensity values
on two "hot pixels" (noted *) have been cut.}
\label{f:oh_fig}
\end{figure}

We observed OH absorption lines around 3080 \AA\ from the 
A-X~(0-0) transition band 
towards HD~34078 with the Canadian-France-Hawaiian
telescope. The observations were performed in 2001, October 4-7. 
We used the GECKO spectrometer, in the UV f/4 Coud\'e train at a
nominal resolving power of R = 120~000. The CFHT EEV1 CCD was used as
the detector. The spectrum was cross dispersed by a
grism and OH lines appear in the 18th order. An integration time of 14 
hours provided a S/N ratio of 36 per 35~m\AA\ pixel. Data were reduced 
in a standard way using the IRAF package. The three OH lines expected
in the range covered are clearly detected. An extract is presented in
Fig. \ref{f:oh_fig} and equivalent widths are given
in Table \ref{tab.oh}. The inferred OH column density towards HD~34078
is N(OH)~=~3.5 $\pm$ 0.5 $\times10^{13}$ \cmd.

\begin{table*}[h!]
\begin{center}
\label{tab.oh}
\caption{Equivalent widths and column densities derived from the OH 
transitions observed towards HD~34078. Wavelengths and oscillator 
strengths are taken from Felenbok and Roueff (\cite{felenbok}).}
\begin{tabular}{cccc}
\hline
\hline
                          & \multicolumn{3}{c}{ Transitions (0-0) X-A} \\ 
$\lambda_{\textrm{air}}$  &  3072.0105 + 3072.0637         & 3078.4399  + 3078.4720         & 3081.6645  \\
Transition                & R$_1$(3/2) + $^R$Q$_{21}$(3/2) & Q$_1$(3/2) + $^Q$P$_{21}$(3/2) & P$_1$(3/2) \\
f                         & 4.55(-4)                       & 1.05(-3)                       & 6.48(-4)   \\ 
W (m\AA)                        & $\le$ 1.0                  & 1.72 $\pm$ 0.27           & 0.86 $\pm$ 0.21\\
N (cm$^{-2}$)             & $\le$ 2.6 $\times10^{13}$      & 1.9 $\pm$ 0.3 $\times10^{13}$& 1.6 $\pm$ 0.4 $\times10^{13}$\\
\hline
\end{tabular}
\end{center}
\end{table*}  

\section{Properties of the gas along the line of sight}
In this section, we study physical properties such as excitation,
temperature and kinematics. Small scale structure is considered in
the next section. We first discuss parameter values as inferred  
from the analysis of absorption lines measured in the first spectrum 
and then present a model attempting to account for all known  
observational constraints.

\subsection{Observational results and comparison 
to other lines of sight}

\begin{figure}
\centerline{\includegraphics[width=6.4cm,angle=-90]{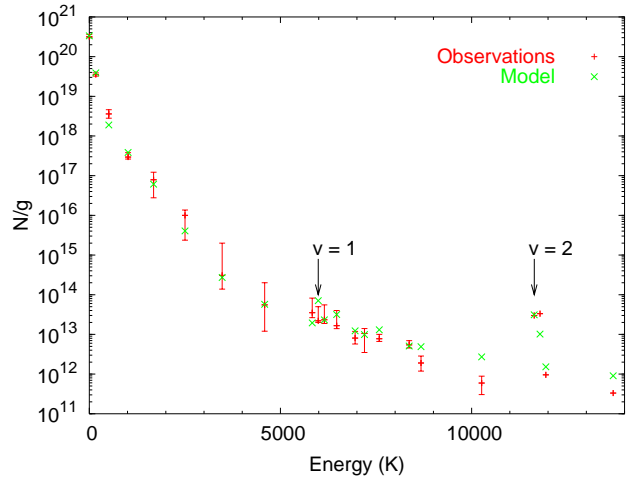}}
\caption{Observed \hdeux\ excitation diagram (filled squares with 3$\sigma$
error bars shown; above 11000~K, observed values are upper limits)) 
and predictions of the model. The energies of the
lowest vibrationally excited v=1, 2 \hdeux\ levels are indicated.}
\label{f:diagexTOT}
\end{figure}

In Fig. \ref{f:diagexTOT} we give the excitation diagram corresponding to the
estimates or upper limits listed in Table \ref{nh2}. 

As observed for other lines of sight, there is a clear flattening as 
excitation energy increases. The J = 0 and 1 values provide a direct 
measurement of the gas temperature, T = 77 K (71 - 83 K).
 The slope inferred from the J = 3 to 7 values correspond to 
T$_{exc} \simeq$ 360 K, whereas 
the detections up to energies of 10$^4$ K are roughly fitted by 
T$_{exc} \simeq$ 1200 K. No J parity effect is noticeable.

\begin{figure}
\centerline{\includegraphics[width=6.4cm,angle=-90]{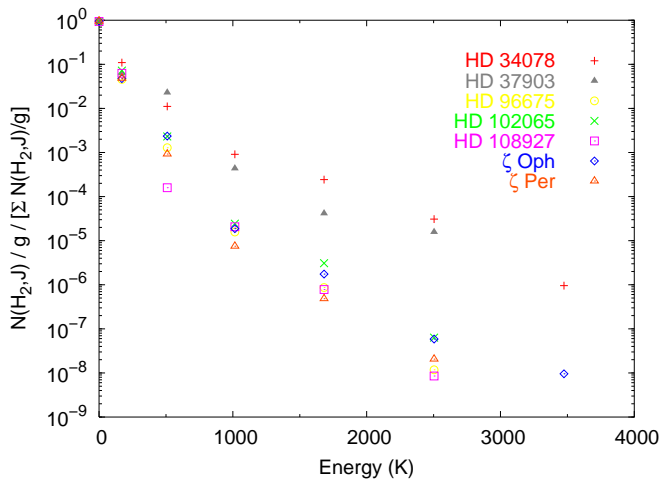}}
\caption{Comparison of the \hdeux\ excitation diagrams observed for 
the lines of sight towards HD~34078, HD~37903, HD~96675, HD~102065, 
HD~108927, $\zeta$
Oph and $\zeta$ Per. Column densities for the $J=0$ to 6 levels have 
been normalised to the total \hdeux\ column density for each line of 
sight. For HD~102065, the N(J) values correspond to a b value of 3 \kms.}
\label{f:diagexCOMP}
\end{figure}

In order to assess whether the \hdeux\ excitation towards
HD~34078 differs from the one determined on other lines of sight 
already at low J (and not only at high excitation energies), we present in
Fig. \ref{f:diagexCOMP} a compilation of results for \hdeux\ levels
from J = 0 to 6 for HD~96675, HD~102065, HD~108927 
(Gry et al. \cite{cecile}), $\zeta$ Oph 
(Morton \cite{morton75}) and $\zeta$ Per (Snow \cite{snow77}; the
Copernicus data have been reanalysed using new oscillator strength
values as described in Le Petit et al. \cite{lepetit04}). 
The first three lines of sight have been chosen because the gas probed 
is known to be far away from the star.

We also include in Fig. \ref{f:diagexCOMP} results obtained for 
HD~37903, the illuminating star of NGC~2023 in Orion. In HST/STIS 
spectra of that B1.5~V star, Meyer et al. (\cite{meyer01}) detect a
large amount of highly excited \hdeux, which is consistent with UV 
fluorescent excitation of dense molecular gas located close to 
HD~37903. The data shown in Fig. \ref{f:diagexCOMP} for J $\le$ 6 are 
from ORFEUS observations (Lee et al. \cite{lee02}).

For the purpose of comparison, column densities have been normalised 
to the total \hdeux\ column density appropriate for each line of sight. 
It is clear in Fig. \ref{f:diagexCOMP} that for the two 
lines of sight in which strong absorption from excited \hdeux\ is 
detected (from J $\ge$ 8), HD~34078 and HD~37903, the excitation excess 
is already apparent at J=2 with 
respect to other, more ``standard'', lines of sight.
Beyond J=3, the excess attains two orders of magnitude. 
In contrast, the relative amount in the J=1
level (corresponding to T $\approx$ 77~K towards HD~34078) is quite 
consistent with other measurements. 

Apart from the detections presented in this paper, several other 
molecular species have been observed towards HD~34078 which can be
used to constrain a physical/chemical model of
the gas: \c2\ (Federman et al. \cite{federman94}), CH, \chp, CN (Rollinde et
al. \cite{emmanuel}) and CO (Mac Lachlan \& Nandy \cite{maclac}). The
corresponding column densities are summarized in Table \ref{tabobserv}.
It appears that, on one hand, column densities of molecular species as
well as the \hdeux\ temperature, T$_{01}$, are qualitatively similar to
those found on other lines of sight and, on the other hand, the \hdeux\ 
excitation above J~=~2 is unusually high compared to that in
standard diffuse ISM. 
This suggests the presence of two distinct
components along the line of sight: i) "quiescent" gas within a 
translucent cloud, responsible for the absorption due to \hdeux\ J~=~0 and 1 
and to other molecular species and ii) gas located close to HD~34078 
dominating absorption from \hdeux\ at J $>$ 2. 

The HI/\hdeux\ ratio only provides a lower bound for comparison with
model predictions since diffuse HI not associated to molecular gas is
likely to be present along the line of sight. Adopting 
the N(HI) value derived from our FUSE observations, the resulting 
total hydrogen column
density along the line of sight, N(H) = N(HI) + 2 N(\hdeux), is
$3.3\cdot10^{21}$ cm$^{-2}$. Given the E(B-V) value quoted in 
Table \ref{tab.param}, the implied N(H)/E(B-V) ratio is consistent 
with the value measured by Bohlin et al. (\cite{bohlin}).

\subsection{Modelling}

To investigate the ability of our model to reproduce
observed column densities we use a recent version of the PDR
model of Le Bourlot et al. (\cite{lebourlot93}) as well as the MHD shock
model of Flower \& Pineau des For\^ets (\cite{flower98}), updated for
diffuse clouds and including explicit rotational excitation of \hdeux. As
described in Le Petit et al. (\cite{franck}), the PDR model features a
plane-parallel slab of gas and dust illuminated on one side by a UV
radiation field. The radiative transfer as well as the thermal balance are
calculated (the temperature can also be fixed, if derived from 
observations). The J dependent photodissociation of molecules such as H$_2$ 
and HD as well as their excitation in ro-vibrational levels due to radiative
excitation, decay and collisional excitation are computed. About 100
chemical species linked by a network of 1400 chemical reactions have been
taken into account. The elemental abundances and dust size distribution used 
here are those given in Le Petit et al. (\cite{franck}). 
Note that while the extinction curve effectively 
measured towards HD~34078 is used to describe the attenuation 
of the UV field, the dust size distribution adopted is consistent 
with the average Galactic extinction curve.
The main properties of the line of sight (extinction curve and R$_\textrm{v}$
value) are given in Table \ref{tab.param}. The density and intensity
of the incident radiation field in each of the two components are
adjusted to reproduce the observed H$_2$ excitation as well as the 
amount of other molecules. Column densities are computed through
integration of the local abundances given by the model, assuming 
a face-on geometry.

\subsubsection{The transluscent cloud}

For the translucent cloud we assume an incident radiation field 
equal to that in the solar neighborhood as estimated by Draine
(\cite{draine78}). An isothermal model is adopted at T~=~77~K, the 
kinetic temperature from the two first \hdeux\ levels. The gas density is
well constrained by absorption from the first three CI levels. 
Trying several values for n$_H$, models with densities of 600
cm$^{-3}$ and below appear to underestimate N(CI) whereas those with 
n$_H~\ge~800$ cm$^{-3}$ overestimate N(CI$^*$). We adopt n$_H$~=~700
cm$^{-3}$ as the best compromise. With these parameters, integrating 
abundances to match the observed N(H$_2$, J=0) and N(H$_2$, J=1), we find 
that column densities of all other molecules (excepted
CH$^+$ and OH discussed below) are well reproduced given the measurement
uncertainties and the small number of free parameters in the model
(cf. Table \ref{tabobserv}). 
The visual extinction corresponding to this component is
A$_{\textrm{v}}$ = 0.8. 

To account for the \chp\ abundance we add to the PDR component 
a C-shock with a velocity  v = 25 \kms, a pre-shock 
density of 20 cm$^{-3}$ and a magnetic field of 7 $\mu$G. The
shock contributes only slightly to the excitation of H$_2$ for levels
J = 5 to J = 7 (Fig. \ref{f:diagexMOD}). The thickness of the shocked 
layer corresponds to A$_{\textrm{v}} \approx $ 0.1 and gives a negligible
contribution for other species, OH excepted. In this latter case, the 
contribution of the MHD shock is too large by a factor of about 7 
(Table \ref{tabobserv}). 
The maximum temperature of neutral species reached in the shock is 
4000~K. A single shock has been used here; vortices or a collection of
several weaker shocks could have been considered as well 
to represent the dissipation of kinetic energy 
along the whole line of sight (cf. Gredel et
al. \cite{gredel}), which might help to reduce the 
discrepancy between the observed and model N(OH) values. However,
since the C-shock required to account for N(\chp) appears to 
play no important role for the \hdeux\ excitation, no further 
attempt to fit both the observed OH and \chp\ values was made
(OH observations are relatively scarce due to the
wavelength range of its absorption lines and thus little
is known about correlations between OH and \chp). \\

\subsubsection{The hot PDR around HD~34078}

The contribution of material surrounding HD~34078 is
strongly suggested not only by the anomalous excitation diagram but 
also by the IRAS detection of a far infrared excess
associated with this star by van Buren et al. (\cite{vanbur2}). 
According to these authors, 
such an emission is indicative of a bow shock located around the star, 
at the interface between the stellar wind and the ambient interstellar
medium, where the material is strongly compressed. In this location, 
dust grains are exposed to an intense UV field and emit FIR radiation, 
accounting for the excess observed in IRAS data. We shall assume in
the following that a scenario like the one proposed by Mac Low 
et al. (\cite{maclow}) to describe cometary compact HII regions also
applies in our case and that \hdeux\ is present in a thin layer 
within the bow shock (cf their Fig. 2). This is not obvious here since
the ambient medium is not necessarily dense enough to contain \hdeux\ 
molecules (this point is discussed in Sect. 4.2.3). Below, we 
consider this bow shock as a PDR and show that excitation of the \hdeux\ 
molecules by UV photons can explain the observed excitation. 

\begin{figure}
\centerline{\includegraphics[width=6.6cm,angle=-90]{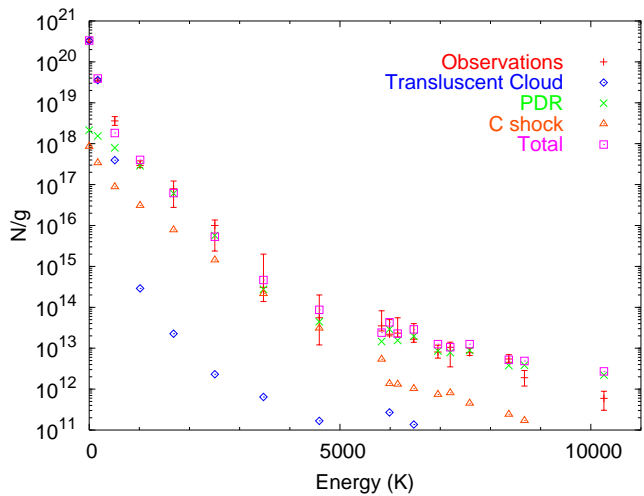}}
\caption{Observed \hdeux\ excitation diagram and \hdeux\ column
densities for each of the three components in the model.}
\label{f:diagexMOD}
\end{figure}

Since we have no observational constraints
on the temperature for this component, we now consider models in which
the thermal balance equation is solved. The density n and the intensity of the
incident radiation field $\chi$ expressed in Draine's units are varied
so as to reproduce the \hdeux\ excitation diagram. We find that adopting 
n$_H$= 10$^4$ cm$^{-3}$, $\chi~=~10^4$ and integrating abundances
up to A$_{\textrm{v}}$ = 1.0, we get a remarkably good fit given the 
limited number of free parameters, as illustrated in 
Figs. \ref{f:diagexTOT} and \ref{f:diagexMOD}. \hdeux\
column densities for levels J~=~3 and above are correctly 
reproduced, indicating that the hot PDR naturally accounts for both the 
overpopulation measured for J $\ge$ 3 and the high energy tail 
(beyond J~=~7) not seen in most lines of sight (let us recall that
N values are well constrained by observations both for J~=~3 and 
J $\ge$ 8 since in these two limits W no longer depends on b). 
Then, the "standard" 
\hdeux\ J = 2 to 7 population as shown in Fig. \ref{f:diagexCOMP} is
masked on this particular line of sight by the hot PDR component.  
Most of the observed atomic hydrogen comes from this component. 
Because of the high radiation field, column densities of molecules
like CO, CH, OH etc are negligible as compared to those of the
translucent component due to large photodissociation rates 
(cf. table \ref{tabobserv}).

By varying n and $\chi$ around the best fit values, we find that 
the density is well constrained by the model: with $\chi~=~10^4$, 
the amount of high excitation H$_2$ can be reproduced only with 
n$_H$ between 5000 and 5 $10^4$ cm$^{-3}$. 
On the other hand, the intensity of the radiation field is poorly 
constrained. For a density n$_H$~=~10$^4$ cm$^{-3}$ it is possible 
to reproduce the \hdeux\ excitation diagram without exceeding the 
observed N(H I) and N(H$_2$) with $\chi$ between 500 and 
10$^4$. We thus find that a two-component model (complemented by a C-shock to
account for \chp\ production within the translucent cloud) 
successfully reproduces the main
characteristics of the gas along the line of sight towards HD~34078.

A comparison of the excitation diagrams obtained for HD~34078 and
HD~37903 (Fig. 6, Fig. 7 and Fig. 2 from Meyer et al. \cite{meyer01})
indicates that they are relatively similar. Moreover, it is noteworthy that
the amounts of \hdeux\ and dust are comparable on these two lines of sight:
N(\hdeux) = 6.4 10$^{20}$ and 7.8 10$^{20}$ \cmd\ and E(B-V) = 0.52 and 0.35
for HD~34078 and HD~37903 respectively (the spectral types are also close to
each other: O9.5V and B1.5V). It is then not surprising that when modelling
the high energy part of the HD~37903 diagram, Meyer et al. (\cite{meyer01})
derive values for the radiation field and gas density consistent with our
estimate for the hot PDR component around HD~34078. In our picture, the
markedly different proper velocities of these two stars (only HD~34078 is a
runaway) could just correspond to distinct mechanisms responsible for the
presence of high density gas close to each star.

\subsubsection{Discussion}

As already noted, we adopted in the above model the true 
HD~34078 extinction curve but
a dust size distribution corresponding to the average Galactic extinction 
curve. This lack of consistency could have significant effects on model 
results. To estimate the uncertainties related to the (unknown) dust 
size distribution, we ran additional models using the average Galactic 
extinction curve (including the appropriate R$_\textrm{v}$ value) instead 
of the HD~34078 one and imposing similar values as those given in 
Table \ref{tabobserv} for the \hdeux\ column density of the translucent 
cloud and hot PDR components. We find that the
results of both models remain within the observational uncertainties; in the 
\hdeux\ excitation diagram, the maximum difference in N(\hdeux, v, J) is 
less than 25\% and similarly small relative variations are seen for the 
abundances of species like CI, CH, CN (the variation attains 50\% 
for CO). We then conclude that our model results are not strongly 
dependent on the adopted dust size distribution and that the good 
agreement with observations is fully significant.

In order to better assess the reliability of the hot PDR component, 
one can estimate from dynamical arguments the distance of the
bow shock to the star and infer the corresponding radiation field
intensity. Using Eq.(2) from van Buren et al.
(\cite{vanbur1}), adopting an average density of the
ambient ISM of $n_a \simeq$ 1 cm$^{-3}$ and a star velocity 
of 100 \kms, 
we get a distance of $d_s \simeq$ 0.5 pc. Considering the observed
flux of HD~34078 at $\lambda \simeq$ 1100 \AA\ (10$^{-11}$ 
erg.cm$^{-2}$.s$^{-1}$.\AA$^{-1}$) and correcting for the
dust attenuation implied by E(B-V) = 0.5 and the observed extinction
curve ($\tau_{1100} = 6.3$), we find that the distance $d_{rad}$ at 
which the intensity of the radiation field of the star is equal to 
10$^4$ times the ISRF is 0.2 pc, approximately the estimated 
distance of the shock to HD~34078. 
This good agreement gives coherence to our model. 
An additional test would consist in estimating the density 
of the material located in the bow shock, 
but unfortunately this is strongly dependent on the magnetic field and
the shock velocity (both unknown).

Although the spectral resolution is limited, one can also use the
absorbing gas kinematics to test our model. In the latter, all \hdeux\
absorptions from levels above the J=2 one arise from the same
component, the hot PDR. These lines should therefore be characterized
by a common velocity. To search for a possible Doppler shift
among J $\ge$ 3 lines we selected spectral regions with good S/N 
ratio and displaying J = 3, 4, 5 lines with closeby J = 9 absorptions. 
Similarly, around 1059 \AA, a group of 7 \hdeux\ lines from v = 0, J = 4,
6, 7, 8 and v = 1, J = 3, 4 is present. In all these cases, we find
that the r.m.s. scatter in wavelength shifts is no larger than
0.005 \AA, which implies a velocity difference smaller than 1.5 \kms\ (or
about 5 \kms\ at 3$\sigma$).

Conversely, since the translucent and hot PDR components are
unrelated in our model, their velocity distributions should differ
both in the average and width, unless
they happen to coincide by chance. J = 0 and 1 absorptions are too
broad to provide any valuable velocity estimate. We can use instead 
HD lines which trace the same component. Near the 1031 and 1054
HD J = 0 absorptions, we find a few narrow \hdeux\ absorptions 
(from the hot PDR component) and by comparing measured wavelengths to
the laboratory values, we find no significant shift between the two
components to a similar limit as above (5 \kms\ at 3$\sigma$). 
The test is however not so
constraining because the number of measurements is limited 
and the accuracy of the latter is difficult to assess because i) the
continuum is uncertain and ii) unrecognized blends may be present.
Regarding velocity dispersions, one would expect a priori a larger
value for disturbed gas near HD~34078 (i.e. the hot PDR
component) than for the foreground translucent cloud. 
High resolution observations of CH absorption can be used to probe the
latter, which indicates b $\approx$ 3 \kms\ (Allen \cite{allen94}, 
Rollinde et al. \cite{emmanuel}). As discussed in Sect. 2.2, there is 
no clear evidence for a significantly larger dispersion in high 
excitation gas. Thus, tests involving kinematics do not provide
additional support for our two-component model. 

The similarity of the velocity and velocity dispersions of the two
components suggests that perhaps the latter are not 
unrelated. The suggestion made by Herbig (\cite{herbig58}) is 
interesting in this
regard. Based on a comparison of H$\alpha$ and visible continuum
images, he proposed that HD~34078 has recently encountered new cloud
material. In such a scenario, one can imagine that the density of the
ambient medium, $n_{a}$, is about 10-20 \cmt\ instead of 1 \cmt. A
first consequence involves the origin of the excited \hdeux. With $n_{a}$
$\simeq$ 1 cm$^{-3}$, \hdeux\  molecules would not preexist and 
would need to be
formed in the bow shock; we estimate that at n $\simeq$ 10$^4$ cm$^{-3}$,
10$^5$ yrs are required while it takes no longer than 10$^4$ yrs 
(a few 0.2 pc/100 \kms) for the compressed material to drift from the
head of the shock to the sightline (after 10$^4$ yrs, the fraction of
molecular \hdeux\ is less than 10\% of its steady-state value). 
In contrast, if 
$n_{a}$ $\simeq$ 20 cm$^{-3}$, \hdeux\ can be supposed to be already in 
place in the ambient medium. Moreover, applying mass conservation 
for the swept gas leads to a predicted \hdeux\ column density through the
hot PDR which is consistent with the observed one. Indeed, one gets 
the relation $ \pi R^2 . n_a= 2 \pi . R . \Delta R . n_{PDR}$ or
$n_a . R = 2 . \Delta R . n_{PDR} = 2 N_{PDR}$ where R is the radius of the
cylinder containing the swept material (which we assume to be
comparable to d$_s$), $\Delta R$ the thickness of the PDR, $n_{PDR}$ 
and $N_{PDR}$ the density and column density throughout the compressed
layer. We find that $N_{PDR}$ is of the order of $10^{19}$ \cmd, 
comparable to the column density estimate for the hot PDR drawn 
from our best fit model (Table \ref{tabobserv}). 
Finally, in a denser ambient medium the shock will 
occur closer to the star ($d_s$ goes as $n_{a}^{-0.5}$) and 
with $n_{a} \simeq$ 20 cm$^{-3}$, $d_s \simeq$ 0.1 pc 
(instead of 0.5 pc for 1 cm$^{-3}$), quite comparable to $d_{rad}$. 
Thus, a relatively high density for the ambient medium around HD~34078, 
as suggested by Herbig (\cite{herbig58}), appears to be well consistent 
with the modelling of the \hdeux\ excitation.

One question that remains open is the relation between the translucent
component containing the bulk of the \hdeux\ (J~=~0, 1) and the hot PDR
material. The low temperature (77~K) of the \hdeux\ gas, the low velocity
dispersion as estimated from CH and the presence of many molecules that might 
be photodissociated if located too close to HD~34078 suggest that 
the translucent cloud is not associated to the star. In contrast,
the similarity of velocities and velocity dispersions favors a picture
in which the two components have a common origin. Below (Sect. 6), we
propose a few observations which should help to clarify the relation 
between HD~34078 and the bulk of the molecular gas.

\begin{table*}[H!]
\caption{Observed and computed column densities in cm$^{-2}$.}
\label{tabobserv}
\begin{tabular}{|l|ccc|cccc|}
  \cline{2-8}
  \cline{2-8}
\multicolumn{1}{c}{}& \multicolumn{3}{|c}{Observations}                             & \multicolumn{4}{c|}{Modelling} \\
  \cline{1-8}		   
                   & Observations       & Min. Val.          &
Max. Val.          & Transluscent Cloud & C-shock       & hot PDR & Total \\
H$^{(1)}$          & 2.7$\cdot 10^{21}$ & 2.2$\cdot 10^{21}$ & 3.4$\cdot 10^{21}$ & 2.6$\cdot 10^{19}$ & 1.2$\cdot 10^{19}$& 1.7$\cdot 10^{21}$ & 1.7$\cdot 10^{21}$ \\
H$_2$$^{(2)}$      & 6.4$\cdot 10^{20}$ & 6.0$\cdot 10^{20}$ & 6.9$\cdot 10^{20}$ & 6.5$\cdot 10^{20}$ & 5.1$\cdot 10^{18}$ & 2.7$\cdot 10^{19}$ & 6.8$\cdot 10^{20}$ \\
A$_{\textrm{v}}$   & 1.8                &                    &                    & 0.8                & 0.1                & 1.0              & 1.9      \\
HD$^{(2)}$         &                    & $10^{15}$          &
$10^{16}$          & 8.5$\cdot 10^{15}$ & --- & 5.4$\cdot 10^{13}$ & 8.5$\cdot 10^{15}$ \\
CH$^{(3)}$         & 9.5$\cdot 10^{13}$ & 9.3$\cdot 10^{13}$ & 1.1$\cdot 10^{14}$ & 5.0$\cdot 10^{13}$ & 4.1$\cdot 10^{12}$  & 7.4$\cdot 10^{9}$ & 5.4$\cdot 10^{13}$ \\
CH$^+$$^{(3)}$     & 6.5$\cdot 10^{13}$ & 5.8$\cdot 10^{13}$ & 7.1$\cdot 10^{13}$ & 1.9$\cdot 10^{10}$ & 6.0$\cdot 10^{13}$ & 5.0$\cdot 10^{11}$ & 6.0$\cdot 10^{13}$ \\
C$_2$$^{(4)}$      & 5.8$\cdot 10^{13}$ & ---                & ---                & 2.3$\cdot 10^{13}$ & 2.9$\cdot 10^{10}$  & 1.4$\cdot 10^{7}$ & 2.3$\cdot 10^{13}$ \\
C$_3$$^{(5)}$      & 2.2$\cdot 10^{12}$ & 1.7$\cdot 10^{12}$ & 2.7$\cdot 10^{12}$ & 7.2$\cdot 10^{11}$ & ---  & 1.3$\cdot 10^{3}$       & 7.2$\cdot 10^{11}$ \\
OH$^{(2)}$         & 3.5$\cdot 10^{13}$ & 1.4$\cdot 10^{13}$ & 5.6$\cdot 10^{13}$ & 1.4$\cdot 10^{13}$ & 2.5$\cdot 10^{14}$ & 5.9$\cdot 10^{11}$ & 2.6$\cdot 10^{14}$ \\
CN$^{(4)}$         & 3.3$\cdot 10^{12}$ & ---                & ---                & 1.6$\cdot 10^{12}$ & 5.8$\cdot 10^{11}$  & 4.8$\cdot 10^{8}$ & 2.2$\cdot 10^{12}$ \\
CO$^{(6)}$         & 5.7$\cdot 10^{14}$ & 4.6$\cdot 10^{14}$ & 7.2$\cdot 10^{14}$ & 1.1$\cdot 10^{15}$ & 3.0$\cdot 10^{13}$ & 1.1$\cdot 10^{11}$ & 1.1$\cdot 10^{15}$ \\
CI$^{(2)}$       & 9.4$\cdot 10^{15}$ & 3.6$\cdot 10^{15}$ & 1.7$\cdot10^{16}$ & 3.5$\cdot 10^{15}$ & --- &  2.4$\cdot10^{12}$ & 3.5$\cdot 10^{15}$ \\
CI$^{* (2)}$       & 5.8$\cdot 10^{15}$ & 1.6$\cdot 10^{15}$ & 5.8$\cdot 10^{15}$ & 5.7$\cdot 10^{15}$ & --- & 6.9$\cdot 10^{12}$ & 5.7$\cdot 10^{15}$ \\
CI$^{** (2)}$     & 2.2$\cdot 10^{15}$ & 1.1$\cdot 10^{15}$ & 4.0$\cdot 10^{15}$ & 2.8$\cdot 10^{15}$ & --- & 1.0$\cdot 10^{13}$   & 2.8$\cdot 10^{15}$ \\
\                  & & & & & & & \\
\hline
\end{tabular}
\newline
{\small Dash means an unknown value.\\
Last column (Total) corresponds to the sum of the components in
the model. \\
References for the observations are:\\
(1) This work and Mc Lachlan \& Nandy (1984). \\
(2) This work. \\
(3) Rollinde et al. (2003). \\
(4) Federman et al. (1994), Rollinde et al. (2003)\\
(5) Oka et al. (2003).\\
(6) Mc Lachlan \& Nandy (1984)}
\end{table*}

\section{Absorption line variations and small scale 
structure towards HD~34078}

The original purpose of our project was to probe the structure of
quiescent molecular gas. The discovery of high excitation \hdeux, making
the line of sight towards HD~34078 somewhat peculiar, could have been 
embarrassing.
However, the modelling work described above indicates that 
absorption from the J = 0 and 1 \hdeux\ levels (which comprise by far most of
the gas) traces material unperturbed by HD~34078. In the following, 
we first search for variations in the damped line profiles from these
two levels, which can be used to investigate in a direct and efficient 
way column density variations, regardless of any
assumption on the velocity distribution. The five spectra
available span a time interval of 2.7 yr, which corresponds to
linear scales of about 50 AU in the foreground cloud.

In our model, lines from all other \hdeux\ levels originate essentially
from gas surrounding HD~34078. Since the compressed material is
continuously flowing across the line of sight as the star moves, 
the time behavior of these lines should tell us whether this flow 
is regular or chaotic. Further, if there exist density fluctuations 
in this gas or changes in the UV field, this should also result
in variations of absorption lines arising from the hot PDR component. 
In order to select those \hdeux\ levels that display the highest 
sensitivity to changes in n or $\chi$, we use the model presented in 
Sect. 4. We vary the density, n, and radiation field, $\chi$, 
(separately) by a factor of 2, and find that the sensitivity to changes in 
$\chi$ is small for all levels (30 \% at maximum for J = 4 - 5), while 
the response to similar variations in n is much larger with a peak at 
J = 6. N(J = 6) varies by a factor of 3.4, and beyond J = 8 the 
induced variation in N(v, J) is around 50\%, relatively independent of 
the level. Unfortunately, absorptions from levels J = 5, 6, 7, 8 are 
heavily saturated and the observed profiles or W values are weakly 
dependent on N(J) (these lines can be used instead to assess the stability 
of the instrument). We shall then examine the good S/N lines with intermediate 
opacity ($\tau \approx 1 - 3$) arising from high excitation levels 
(J $\ge$ 9).

In Boiss\'e et al. (\cite{patrick01}), we presented a report based on the 
first three spectra, among which no significant changes were noted, except a
possible variation of lines from highly excited \hdeux. 

\subsection{Sensitivity of absorption lines to column density variations}
The sensitivity of an absorption line profile to small column density 
variations can be estimated from the ratio 
$Q \equiv ({{\delta} I/I_c})/({\delta {\rm N}/{\rm N}})$, where $I$ is the
{\it observed} profile and $I_c$ the continuum. 
Considering ideal Voigt profiles observed at a given
finite resolution, Rollinde (\cite{these_emmanuel}) has shown that a
good sensitivity
is achieved in two cases: i) fully resolved lines (e.g. damped \hdeux\
absorption from low J levels) and ii) marginally thin lines (central
opacity, $\tau_0 \approx 1$). 

\begin{figure}[!h]
\centerline{\includegraphics[width=6.4cm,angle=-90]{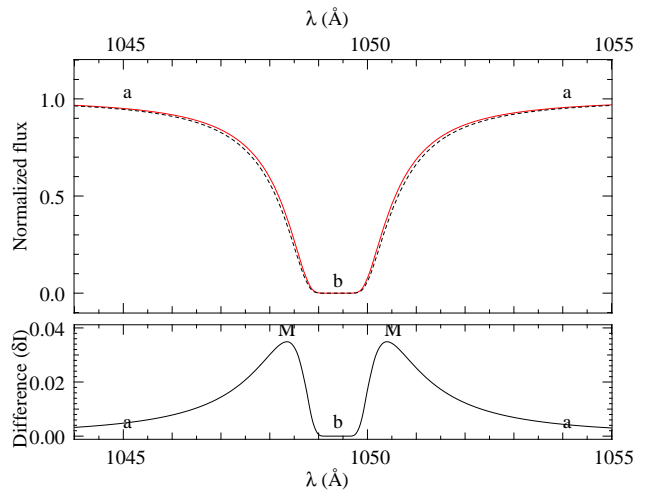}}
\caption{Profile variation for the \hdeux\ Lyman (4-0) R(0) 1049.37 
\AA\ transition. {\em Top panel}: The red line corresponds to 
$\log({\rm N})$ =
20.6. The dashed profile is obtained by varying N by + 10\%.
{\em Lower panel} : difference between the two profiles.
In both panels, regions noted ({\bf a} and {\bf b}) are subject to
little variation which implies the presence of a maximum ({\bf M}) in
between.}
\label{f:variation_l}
\end{figure}

Consider an unblended and fully resolved damped line. If the column
density increases the profile broadens, which results in a decrease of
$I$/$I_c$ at any given wavelength, except in the line core and outside the 
wings where $I$ = $I_c$ (Fig. \ref{f:variation_l}). 
It can be shown that the Q parameter reaches its 
maximum, Q$_{max}$ = 0.357, on each wing of the profile at wavelengths where 
the optical depth $\tau_0$ equals 1 and $I$/$I_c$ = 0.37 
(Rollinde \cite{these_emmanuel}). 
For a change of N by 10\%, the expected variation in $I$/$I_c$ is
0.035 (well above the noise level in our FUSE
spectra).
The corresponding broadening of the profile at $\tau_0 = 1$ on each wing is
typically 0.05 \AA\ (a value comparable to the resolution).  

A damped profile corresponds to 
$$I = I_c\, \exp\left(-\left[\frac{6.726 \times 10^{-17}}{c^2}\right]\times \frac{{\rm N}\, f\, \lambda_0^2\, \gamma}{(\lambda_0/\lambda-1)^2}\right)$$
where c is the light velocity (in \kms), f the oscillator strength,
$\gamma$ the radiative rate (in $s^{-1}$), and where $\lambda_0$ and N are
expressed in \AA\ and \cmd\ respectively. Then, if variations are 
observed for one specific line, one can predict exactly the
corresponding variations expected for other damped lines from 
the same level. This gives a powerful method for checking the reality of 
any variation and rule out artefacts of instrumental origin (e.g. due
to distortions in the wavelength calibration).

\begin{figure}[!h]
\centerline{\includegraphics[width=6.4cm,angle=-90]{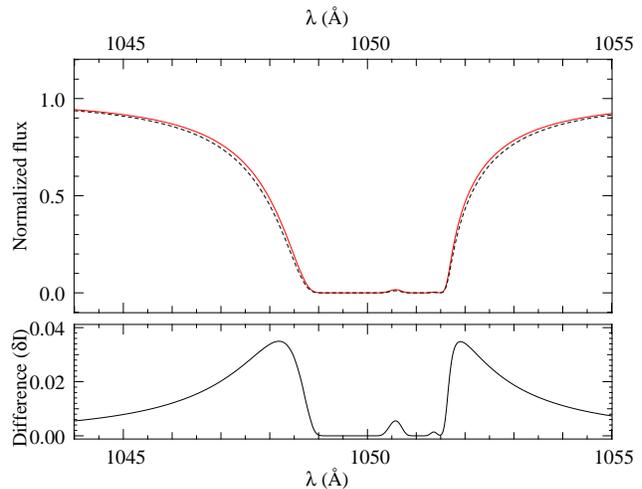}}
\caption{Same as in Fig. \ref{f:variation_l} but for the whole blend of
J = 0, 1, 2 lines ((4-0) Lyman band). The assumed variation is 
$\delta$ N/N = 10 \% for each of these three levels.}
\label{f:variation}
\end{figure}

In practice, the low J \hdeux\ transitions in FUSE spectra appear in "systems",
where one J = 0, two J = 1 and one J = 2 lines are strongly blended 
(e.g. for the (4-0), (3-0) and (2-0) Lyman bands at $\lambda \approx$ 
1050 \AA, 1063 \AA, 1078 \AA\ respectively). To compute the 
expected variation of the whole system profile, we assume that the 
excitation of the gas is uniform
i.e. that the relative variations of N(J = 0), N(J = 1) and N(J = 2) are
identical. Comparison of synthetic profiles corresponding to values of
N(J = 0, 1 and 2) similar to those given by profile fitting of
the FUSE spectra indicates that the blue wing of the J = 0 line and
the red wing of the J = 2 line in each system (the two extreme
transitions of the blend) are relatively little affected by the J = 1
transitions (Fig. \ref{f:variation}). For the 1050 \AA\ \hdeux\ system, 
changes in N(J = 1) alone result in profile variations that would not 
be easy to detect because they appear in the far wings of the blend; 
as a consequence, no useful constraint can be obtained for N(J = 1)
from that system.

\subsection{Comparison of \hdeux\ damped profiles obtained between 
Januray 2000 and October 2002}

To search for variations among the observed spectra one could fit each
system profile (or fit them simultaneously) and then compare the N(J=0, 1,
2) values obtained for each epoch. However, the accuracy of individual N
values would then be strongly limited by our knowledge of the stellar
continuum, which is poor given the large width of the \hdeux\ absorptions
and high number of additional absorption lines present.
A great advantage of {\it reobserving the same star} is that variations can
be searched for through a direct comparison of the spectra themselves.

Before performing this comparison, we need to adjust the wavelength
and flux scales of the two spectra considered. These two tasks are performed
separately for each \hdeux\ system, on an extract of the spectra centered
on the absorption and covering about 15 \AA. Using narrow absorption
lines located on both sides of the broad absorptions, the wavelength
shift to be applied is measured. Similarly, a multiplicative factor 
bringing the adjacent continua of the two spectra to the same level is 
determined. In fact the "adjacent continuum" is not easy to define 
because the various systems overlap, and given our N(\hdeux, J) values 
for J = 0, 1, 2 the entire spectral domain of interest is affected 
by \hdeux\ absorption. Thus, to perform the relative flux normalisation, 
we use regions where \hdeux\ absorption is minimum, about 7 \AA\ away 
from the center of each trough (e.g. near 1043, 1057, 1071 \AA). 
It is to be noted that such a procedure 
is approximate only, because a variation in N(\hdeux) will slightly 
affect the continuum level, and imposing the same continuum value in 
the two spectra to be compared will result in an apparently lower 
variation. However, in our case, the absorption in between two \hdeux\ systems 
is no larger than about 10\%. By applying the inter-normalisation procedure
described above to synthetic spectra with N(\hdeux, J) differing by 
10\%, we find that the apparent reduction in the N variation is
not really significant, given our accuracy. In the following, we
therefore ignore this difficulty.

\begin{figure}[!h]
\centerline{\includegraphics[width=6.4cm,angle=-90]{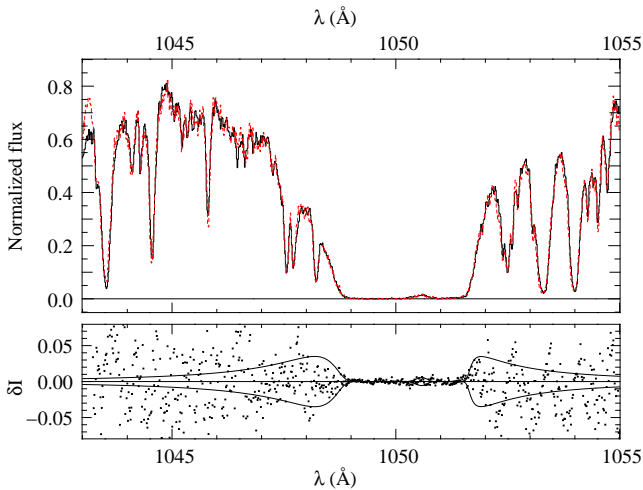}}
\caption{Comparison of the January 2000 and October 2002 
HD~34078 FUSE spectra in the region
comprising the 1050 \AA\ \hdeux\ system. 
{\em Top panel}: Superposition of the normalised 
spectra (black line: Jan. 2000; red line: Oct. 2002), after applying the
corrections discussed in the text. 
{\em Lower panel} : The difference between the two observed normalised
profiles. The solid lines indicate the variation expected
of the \hdeux\ absorption profile for a change $\delta$N/N = $\pm$ 10\% of 
the column density for each of the J~=~0, 1 and 2 levels.}
\label{f:compare}
\end{figure}

These two corrections (involving two values only for a given 
\hdeux\ system and pair of spectra) account 
i) for unknown changes in the zero-point of the FUSE spectra from one epoch
to another (and possibly, for inaccuracies in the FUSE wavelength
calibration) and ii) for stellar flux variations (and/or variations in the
flux calibration). The second correction assumes that no variation in the
spectral shape of the continuum has occurred, which appears consistent with
the data (except blueward of the OVI doublet where impressive
variations of the absorption due to the stellar wind are seen). As shown in
Fig. \ref{f:compare}, the superposition is excellent once the
two corrections are applied, which indicates a very good instrumental 
stability. 

Variations of the \hdeux\ systems among all spectra have been examined (see
Boiss\'e et al. \cite{patrick01} for a comparison of the first and third
spectra). Here, we present results involving 
the first and fifth spectra, which provide the largest timebase. The
comparison has been performed for each of the three 1050, 1063 and 
1078 \AA\ \hdeux\ systems. In Fig. \ref{f:compare} we can see that the
difference near the red or blue end of the through where the largest
variations are expected is no larger than a few percent of the
continuum (windows of width $\simeq$ 0.3 \AA\ are considered) 
and that no systematic trend towards an increase or
decrease of N(\hdeux) is noticeable (the curves shown in the lower
panel are for N(\hdeux) 
changes by $\pm$ 10 \%). For systems other than those around 1050, 
1063 and 1078 \AA, the S/N ratio is too low or the
presence of additional absorption strongly alters the \hdeux\ profile
in the spectral regions where the largest variations are
expected for $I$/$I_c$ ; in particular, we exclude the 1037 \AA\ 
system which is affected by variable OVI absorption from the stellar wind.

Very recently (september 2003), we obtained a sixth FUSE spectrum which
brings the time interval and scale covered to 3.7 yr and 63 AU
respectively. Although these data have not been fully analysed yet, we could
verify that the profiles of the three \hdeux\ systems considered above 
are again completely consistent with those of the earlier epochs.

We thus detect no appreciable change in the damped \hdeux\ 
profiles. To get an upper limit on column density variations, 
we shall use a Maximum Likelihood Estimator (as described in 
Numerical Recipes, Sect. 15.1).
The data correspond to the flux variations along the
profile of the three systems mentioned above, while column 
density variations are the parameters to be determined.
For simplicity, we shall consider that \hdeux\ excitation 
among the J~=~0, 1 and 2 levels is uniform, i.e. constant in time
(clearly, local changes in the excitation would constitute a 
second order effect). In this assumption, the observed profiles
that are blends of absorption lines arising from these three levels
depend only on one single parameter, the total (or relative) N variation.

The {\em likelihood} of $\delta N$ given a flux variation $\delta I$
is a function of the wavelength, written as
 $P(\delta N | \delta I,\lambda)$.
This probability distribution is computed from random 
realisations of synthetic spectra with different column 
densities (using our fit to the data as a reference) to which we add 
white noise with an r.m.s. scatter similar to that measured
in the observed spectrum. Then, we get the combined {\em likelihood} 
function $P(\delta N | \delta I)$ for the set of three \hdeux\ systems
quoted above by computing the normalised product of the three 
individual distributions. The data imply $\delta N/N = 0.02 \pm 0.01$,
which corresponds to 3$\sigma$ lower and upper bounds,  
$ -1 \% < \delta N/N < 5\% $.

Observations of the CH and \chp\ blue lines have suggested possible
correlated variations of N(CH) and N(\chp) (Rollinde et al. \cite{emmanuel}) on
the scale of about one year. One may then wonder in case these fluctuations are
real (additional observations are in progress to check this), whether
they are accompanied by similar N(\hdeux) variations. It turns out that,
for two CH measurements which display the largest variation in W(\CHl)
(+ 6.3\% between January 2000 and February 2001 corresponding to 
+ 9.5\% in N(CH)), there exist nearly simultaneous FUSE observations (to
within a few days). We then examined carefully the 1050 \AA\ \hdeux\
system (which provides the tightest constraints) and searched for any 
hint of an {\it increase} in N(\hdeux) between the
two epochs. No such change can be seen and by comparing synthetic
profiles obtained by slightly varying N(\hdeux, J = 0, 1, 2) about the
best fit values we find that an increase of N(\hdeux) as large as 9.5\% 
is clearly ruled out by our data.

Let us conclude this section by a few comments on some systematic
effects that might affect the analysis described above. For
instance, slight distortions in the wavelength scale could, if large
enough, mimic the kind of signature we are searching for (a shift
of the blue or red end of the \hdeux\ troughs). Local
fluctuations in the continuum level due to varying stellar absorption
lines could also induce apparent \hdeux\ absorption changes. However,
among the three systems considered these effects are found to be
negligible, as indicated by the good agreement between the wavelength
shifts derived from various narrow lines adjacent to each system.
If future observations were to suggest real variations, we note
that the spectral shape of the expected variation could be used to 
recognize these effects; moreover, data from other
detectors (such as LiF2B, which covers in part the same range as LiF1A)
could be used also (although of lower S/N ratio) to rule out
instrumental effects.

\subsection{Lines from highly excited \hdeux}

We now search for variations among the low optical depth lines arising
from high \hdeux\ levels and tracing (in our model) the hot PDR
component around HD~34078.
In order to avoid false detection of variations  
due to instrumental changes (e.g. spectral resolution) or to 
variations in stellar features blended with the \hdeux\ line considered, 
we require that at least 
two lines from the same level display the same behavior. We then focus 
on those levels from which several lines with a high degree of 
significance and intermediate opacity values ($\tau_0 \simeq$ 1 - 5) 
are detected (e.g. J = 9, (v = 1, J = 1, 2, 3)). Occasionally, we
observe apparent variations e.g. for the v = 1 lines at 1052.6 \AA\ 
(J = 1) and 1056.8 \AA\ (J =3) for which a decrease in W by about 
20 \% between epoch 1 and 2 is suggested by the data. However, we
cannot find any \hdeux\ level for which unambiguous evidence for a column
density variation can be obtained. Often, the shape of the continuum
adjacent to absorption lines is not stable (probably because of changes in
the profile of underlying stellar absorption) which greatly complicates 
the comparison of spectra and the estimate of uncertainties on W 
or N variations.
We thus conclude that
the FUSE data exclude the existence of large variations in excited \hdeux\ 
lines (larger than about 30\%). Clearly, HST spectra would be very
helpful to better constrain the temporal behavior of highly excited
\hdeux\ absorption thanks to both a higher resolution and S/N ratio.

\subsection{HI}
\begin{figure}[!h]
\centerline{\includegraphics[width=6.4cm,angle=-90]{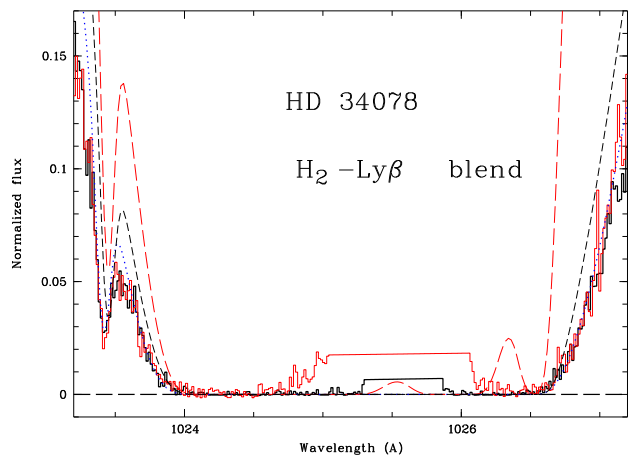}}
\caption{The bottom of the \hdeux\ - Ly$\beta$ trough near 1025 \AA. The
normalized January 2000 (thick black line) and October 2002 (thin red line) are
shown together with synthetic profiles obtained assuming N(HI) = 0 (long-dashed
red line), log(N(HI)) = 21.30 (the value inferred from 1979 IUE Ly$\alpha$ 
data; short-dashed black line) and log(N(HI)) = 21.44 (our best value;
dotted blue line). Note how the latter value improves the fit on both edges
of the trough. The line at 1023.4 \AA\ is from \hdeux\ J = 4. The N(\hdeux)
values for J = 0, 1, 2, 4  used to compute profiles are those given in
Table \ref{nh2}. The emission feature cut out near 1025.5 \AA\ on 
both spectra is terrestrial \lyb.}
\label{f:HI}
\end{figure}

The \lyb\ line offers an interesting opportunity to search for N(HI)
variations. As already mentioned, the N(HI) value obtained from modelling
the \hdeux\ - \lyb\ blend is larger than the estimate drawn from 
IUE observations
(Mc Lachlan \& Nandy \cite{maclac}). In Fig. \ref{f:HI} we show
the spectral region used to determine N(HI) from the Ly$\beta$ - \hdeux\
blend together with synthetic profiles computed with the 1979 and 2000
N(HI) values.

To assess whether the two determinations are
really inconsistent, we have reanalysed the \lya\ data. In the
spectrum that we retrieved from the IUE database there is some weak
residual flux at the bottom of the damped \lya\ profile (while, given
the spectral resolution, the profile should reach the zero level in
the absence of scattered light). Correcting for this offset and
fitting the \lya\ profile again, we get a value - log(N(HI)) = 21.30 -
slightly larger than that given by Mc Lachlan \& Nandy (\cite{maclac}):
log(N(HI)) = 21.24. This revised N(HI) remains significantly lower
than the January 2000 value. Although it is difficult to reach a 
firm conclusion regarding the reality of a change in N(HI) because we 
could not compare the same feature from both epochs and because 
the \lyb\ line is strongly blended with \hdeux\ absorption, our 
analysis suggests that the HI column density towards HD~34078 
has increased by 38\% between 1979 and 2000 (this time interval 
corresponds to a transverse separation of 300 AU
if the HI is assumed to lie at d~=~400pc).  
The corresponding average increase rate for N(HI) would be 
1.8\%/yr, which turns out to be comparable to the rate implied by CH
observations (1.7\%/yr; Rollinde et al. \cite{emmanuel}).

On smaller scales, the ubiquitous structure in the atomic phase
(Heiles \cite{heiles97}) 
should be apparent if our sensitivity to variations in the \lyb\ profile is
good enough (most of the HI is likely unassociated with
HD~34078). Modelling of the profile indicates that the region near
1026.9 \AA\ (rest wavelength), at the red end of the trough, is best
suited to search for N(HI) variation. At longer wavelengths, where
larger variations would be expected, strong lines are present as well
as OVI$\lambda$1031 variable absorption from the O star wind. We
carefully compare the five profiles in this region (after adjusting
the wavelength and flux scales as explained in Sect 4.2 above) and find no
evidence for any variation larger than 10\% (the exact upper limit is
difficult to define but we estimate
that a 10\% variation would unambiguously be seen; 10\% is therefore
equivalent to a 3 $\sigma$ limit).\\

\subsection{Discussion}
The main goal of our work is to study the AU scale structure in the
spatial distribution of \hdeux\ through the temporal behavior of its
absorption lines. Concerning the bulk of molecular gas (i.e. J = 0 and 1
levels) which is believed to lie far from HD~34078, no profile
variation is detected, which implies column density fluctuations lower than
5\% over scales ranging from 5 to 50 AU. We conclude that translucent
clouds display no "ubiquitous" small scale structure in molecular
hydrogen such as that seen for H$_2$CO, OH or in the atomic phase 
through HI 21 cm absorption (Heiles \cite{heiles97}) or NaI optical lines. 
Our sampling remains limited, but since we observed five distinct
positions, from 5 to 50 AU apart, one can reliably conclude that a 
line of sight with N(\hdeux)
$\simeq 3. \times 10^{20}$ \cmd, does not in general intersect
clumps of this length scale with N(\hdeux) larger than 
$1.5 \times 10^{19}$ \cmd.

The observed absence of ubiquitous N(\hdeux) fluctuations in translucent
clouds is an important result in that it
tells us that uniform models can 
legitimately be used to describe chemical and physical processes within 
interstellar gas (which we did above in Sect. 4 !). Structure might be
present at larger scales and it is noteworthy that direct CO emission 
imaging  using millimeter wave interferometers can now probe 
structure at scales as small as 500 AU, i.e. barely one order of 
magnitude larger than those explored here. A few projects of this type 
are in progress; preliminary results indicate some velocity structure
but modest column density variations (Falgarone, private communication).

CH observations have indicated a 20\% increase of N(CH) over the last 12
years (Rollinde et al. \cite{emmanuel}). Since CH and \hdeux\ appear to be well
correlated from one line 
of sight to another, one can search for any such trend in the FUSE
data. The two extreme FUSE observations are separated by 2.7 yr and
show no hint of such a variation. Assuming that the rate of increase is 
constant (about 1.7\%/yr for N(CH); in fact the shape of this
variation is not well constrained: cf Rollinde et al. \cite{emmanuel}), an 
increase of N(\hdeux) by 4.5 \% would be expected if a linear correlation
still holds at small scales. This is just below what we can detect. 
Therefore, the present data do not allow us to check whether the long-term
evolution of CH and \hdeux\ is similar. 

Regarding highly excited \hdeux, the time behavior of its absorption
lines is related, in the frame of the model sketched in Sect. 4, to the 
stability of the gas flow in the shocked gas layer, at the interface 
of the stellar wind and the ambient medium in which HD~34078
is moving. If the latter is not homogeneous in the 1 - 100 AU scale range,
this will induce line variations resulting from fluctuations of the 
amount of gas swept up by the bow shock and then compressed in the
cylindrical layer around HD~34078. Variations due to instabilities 
in the interface could also occur, even if the ambient medium is homogeneous.
The lack of pronounced variations indicates that the medium through
which HD~34078 is moving is smooth and, further, that marked instabilities
do not develop at the interface.  

\section{Summary and Prospects}

The analysis of five HD~34078 FUSE spectra taken between January 2000
and October 2002 leads to the following conlusions:

- the unexpectedly large amount of higly excited \hdeux\ is attributed to 
the presence of molecular gas in a dense layer surrounding
HD~34078 and excited by fluorescence. Detailed modelling of this region 
provides a very good fit to the excitation diagram, with few free
parameters,

- the bulk of the molecular gas (\hdeux\ (J = 0, 1), CH, \chp, CO etc) lies in
a foreground translucent cloud, presumably unrelated to HD~34078. 
Physical conditions (temperature, density, etc) are similar to those 
seen in diffuse molecular gas in general,

- the stability of damped \hdeux\ absorption lines from J = 0, 1 and 2
levels indicate that \hdeux\ molecules are smoothly distributed within
the foreground translucent cloud with column density fluctuations
lower than 5\% over scales ranging from 5 to 50 AU. 

- \hdeux\ lines from highly excited levels also turn out to be 
stable. Again, this points towards a smooth distribution of 
\hdeux\ molecules in the diffuse ambient medium through which HD~34078 is
presently moving and indicates that no instability is
developing in the shocked gas flow around HD~34078.

Let us conclude with a few prospects concerning the study of this 
very interesting
line of sight. As discussed above, our analysis of small scale
structure assumes that the translucent cloud lies far away in front 
of HD~34078 and thus heavily relies on a good understanding of the 
location of the gas responsible for the various absorptions observed. 
It is then important
to find additional support for the two-component model proposed. We
first plan to map the CO emission in the close environment of 
the HD~34078 line of sight. If our assumption is correct, there should
be no relation between the morphology of the observed CO emission
and the position of the O star or the direction of its motion
(essentially northward). 
While comparing the radial velocity (and velocity dispersion) of the 
absorptions arising from each of the two main components, the limited FUSE
resolution prevented us to reach definite conclusions. 
HST/STIS observations would be invaluable in this respect. 
Thanks to a higher resolution and S/N, HST spectra would also allow us
to much better characterise both the kinematical and physical 
properties of the translucent and hot PDR components, through
detection of \hdeux\ lines from higher excitation levels (cf the study by
Meyer et al. \cite{meyer01}) as well as from CI or CO.
Finally, the \lya\ profile would give an accurate N(HI)
determination and then allow us to confirm the reality of the
long-term N(HI) variation.

\begin{acknowledgements}
We thank J. Black and S. Federman for helpful remarks and the referee for 
several constructive comments which contributed to clarifying and 
strengthening the content of this paper. We also thank D. Meyer for 
providing the numerical values of the column densities of vibrationally 
excited \hdeux\ towards HD37903. This work has been done using the profile 
fitting procedure OWENS developed by M. Lemoine and the FUSE French Team 
that we would like to thank for their help. 

\end{acknowledgements}

\end{document}